\documentclass[aps,prd,superscriptaddress,nofootinbib,showpacs]{revtex4-1}

\usepackage{amsmath}
\usepackage{graphicx}
\newcommand{\<}{\langle}
\renewcommand{\>}{\rangle}

\begin{document}

\title{Exploring strange nucleon form factors on the lattice}

\author{Ronald~Babich}
\affiliation{Center for Computational Science, Boston University,
  3 Cummington Street, Boston, MA 02215, USA}
\author{Richard~C.~Brower}
\affiliation{Center for Computational Science, Boston University,
  3 Cummington Street, Boston, MA 02215, USA}
\affiliation{Department of Physics, Boston University,
  590 Commonwealth Avenue, Boston, MA 02215, USA}
\author{Michael~A.~Clark}
\affiliation{Harvard-Smithsonian Center for Astrophysics, 60 Garden Street,
  Cambridge, MA 02138, USA}
\author{George~T.~Fleming}
\affiliation{Department of Physics, Yale University, New Haven, CT 06520, USA}
\author{James~C.~Osborn}
\affiliation{Argonne Leadership Computing Facility, 9700 S. Cass Avenue,
  Argonne, IL 60439, USA}
\author{Claudio~Rebbi}
\affiliation{Center for Computational Science, Boston University,
  3 Cummington Street, Boston, MA 02215, USA}
\affiliation{Department of Physics, Boston University,
  590 Commonwealth Avenue, Boston, MA 02215, USA}
\author{David~Schaich}
\affiliation{Department of Physics, Boston University,
  590 Commonwealth Avenue, Boston, MA 02215, USA}

\date{December 2, 2010}

\begin{abstract}
  We discuss techniques for evaluating sea quark contributions to
  hadronic form factors on the lattice and apply these to an
  exploratory calculation of the strange electromagnetic, axial, and
  scalar form factors of the nucleon.  We employ the Wilson gauge and
  fermion actions on an anisotropic $24^3\times 64$ lattice, probing a
  range of momentum transfer with $Q^2<1\ \mathrm{GeV}^2$.  The
  strange electric and magnetic form factors, $G_E^s(Q^2)$ and
  $G_M^s(Q^2)$, are found to be small and consistent with zero within
  the statistics of our calculation.  The lattice data favor a small
  negative value for the strange axial form factor $G_A^s(Q^2)$ and
  exhibit a strong signal for the bare strange scalar matrix element
  $\langle N|\bar ss|N\rangle_0$.  We discuss the unique systematic
  uncertainties affecting the latter quantity relative to the
  continuum, as well as prospects for improving future determinations
  with Wilson-like fermions.
\end{abstract}


\maketitle

%%%%%%%%%%%%%%%%%%%%%%%%%%%%%%%%%%%%%%%%%%%%%%%%%%%%%%%%%%%%%%%%%%%%%%

\section{\label{intro}Introduction}

The strange quark is the lightest non-valence quark in the nucleon,
and as such provides a unique window into the structure of the proton
and neutron.  Lattice QCD represents at present the only
first-principles predictive method to determine such contributions
directly from the underlying theory of the strong interaction.  The
computational framework for doing this is well established with no
fundamental barriers to success.  Until recently, however, the
calculation of the required quark-line disconnected diagrams was
simply too computationally demanding to provide statistically
significant results with all uncertainties under control.  This is
beginning to change.  With recent algorithmic advances and continued
increases in available computer resources, a new era is dawning where
these effects might be determined with precision well beyond both
experiment and phenomenological estimates.  Here we report on some
recent progress toward this goal.

Strange nucleon form factors represent an attractive test case because
they have also been the subject of a vigorous experimental program.
In particular, a number of collaborations have sought to measure the
strange electric and magnetic form factors via parity-violating
electron scattering~\cite{Beck:2001yx,Ramsey-Musolf:2005rz}, notably
SAMPLE, A4, HAPPEX, and G0.  Recent combined
analyses~\cite{Liu:2007yi,Young:2006jc} find values for $G_E^s(Q^2)$
and $G_M^s(Q^2)$ that are small and consistent with zero in the range
of momenta so far explored. Also of interest is the strange axial form
factor $G_A^s(Q^2)$, to which electron scattering experiments are
relatively insensitive.  At present, the best constraints come from
the two-decades old neutrino scattering data of the E734 experiment at
Brookhaven~\cite{Ahrens:1986xe}.  A recent
analysis~\cite{Pate:2008va}, combining these results with those of
HAPPEX and G0, favors a negative value for $G_A^s(Q^2)$ in the range
$0.45 < Q^2 < 1.0$~GeV$^2$.  These may be compared with the recent
MiniBooNE result~\cite{AguilarArevalo:2010cx}, which is compatible but
at the same time consistent with zero.

A special case is presented by the strange axial form factor at zero
momentum transfer, $G_A^s(0) = \Delta s$, which may be identified with
the strange quark contribution to the spin of the nucleon.  This
quantity is of particular importance, given the role sea quarks are
thought to play in resolving the proton ``spin
crisis''~\cite{Kuhn:2008sy}.  In principle, it is accessible in deep
inelastic scattering, where it is given by the first moment of the
helicity-dependent structure function $\Delta s(x)$.  In practice,
however, determining the first moment requires an extrapolation of the
experimental data to small values of Bjorken $x$, where uncertainties are less
under control.  There is some tension between the two most recent
analyses from HERMES~\cite{Airapetian:2007mh,Airapetian:2008qf}, which
rely on different techniques; the former favors a negative value for
$\Delta s$ while the latter finds a result consistent with zero,
within somewhat larger uncertainties.

Unlike the strange electromagnetic and axial form factors, the strange
scalar form factor $G_S^s(Q^2)$ is not directly accessible to
experiment.  At zero momentum transfer, this quantity corresponds to
the strange scalar matrix element $\langle N|\bar ss|N\rangle$.  Often
considered in relation to the pion-nucleon sigma
term~\cite{Gasser:1990ce}, it is an important parameter in models of
the nucleon.  We also note the pivotal role it plays in the
interpretation of dark matter experiments.  Many models of TeV-scale
physics, including the MSSM, yield a dark matter candidate (e.g.,
neutralino) that scatters from nuclei via Higgs exchange.  The Higgs
is believed to predominantly couple to strange quarks in the nucleon,
since the lighter quarks have proportionally smaller Yukawa couplings
while the heavier quarks are too rare to contribute significantly.  It
follows that the cross section is particularly sensitive to the
strange scalar matrix element, which enters through the parameter
\begin{equation}
f_{Ts} = \frac{m_s \langle N|\bar ss|N\rangle}{M_N}\,.
\end{equation}
As emphasized recently
in~\cite{Bottino:2001dj,Ellis:2005mb,Baltz:2006fm,Ellis:2008hf},
$f_{Ts}$ is poorly known at present and represents the leading
theoretical uncertainty in the interpretation of direct detection
experiments.  A commonly used estimate is that of Nelson and
Kaplan~\cite{Nelson:1987dg,Kaplan:1988ku} (via~\cite{Baltz:2006fm}),
who find $f_{Ts} = 0.36(14)$.  More recent analyses suggest that the
quoted uncertainty may be underestimated~\cite{Bottino:2001dj}, and as
we discuss further in Section~\ref{scalar}, recent lattice
determinations favor a much smaller value.  If these results prove to
be robust, they would imply that the strange quark is in fact not the
dominant contribution, and predicted cross-sections should be
substantially smaller as a result~\cite{Giedt:2009mr}.

\begin{figure}
\begin{center}
\includegraphics*[width=10cm]{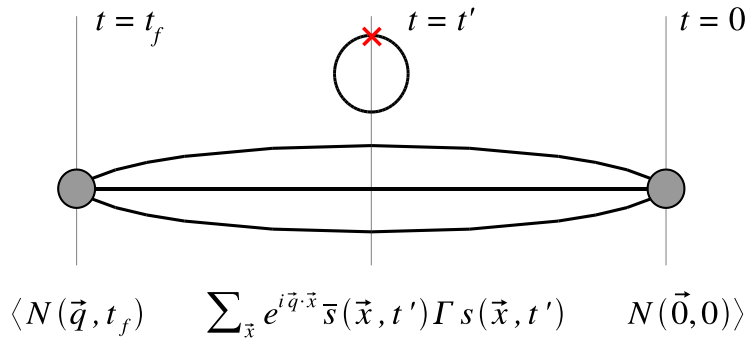}
\caption{\label{fig-disco}Schematic representation of a disconnected
  diagram, giving a strange form factor of the nucleon.  Here $\Gamma$
  is the appropriate gamma insertion for the form factor of interest,
  and $N$ is an interpolating operator for the nucleon.}
\end{center}
\end{figure}

In recent years, there has been a great deal of progress in probing
the structure of the nucleon on the
lattice~\cite{Zanotti:2008zm,Alexandrou:2010cm}.  With few
exceptions, however, such studies have been restricted to the
determination of isovector quantities or otherwise neglect the
contribution of ``disconnected diagrams,'' due to the large cost
associated with their evaluation.  Nucleon matrix elements involving
the strange quark are inherently disconnected, making them an
excellent test case for tackling this challenge.  Such a matrix
element is shown schematically in Figure~\ref{fig-disco}; by
``disconnected,'' we mean that the diagram includes an insertion on a
quark loop that is coupled to the baryon correlator only via the gauge
field.  This requires the calculation of a trace of the quark
propagator over spin, color, and spatial indices.  Since an exact
calculation would require a number of inversions proportional to the
spatial lattice volume, the trace is generally estimated stochastically, which
introduces a new source of statistical error whose reduction is
discussed in Section~\ref{trace}.

There has been a resurgence of interest lately in computing
disconnected form factors on the lattice, building on the pioneering
studies of a decade ago~\cite{Fukugita:1994fh,
  Fukugita:1994ba,Dong:1995rx,Dong:1995ec,Dong:1997xr,Gusken:1998wy,
  Gusken:1999as,Mathur:2000cf,Lewis:2002ix}, which were mainly carried
out in the quenched approximation.  Recent work has included a
determination of the nucleon's strange electromagnetic form
factors~\cite{Doi:2009sq}, preliminary determinations of $\Delta
s$~\cite{Babich:2009rq,Bali:2008sx,Bali:2009hu,Bali:2009dz,Collins:2010gr},
and several studies of the strange scalar matrix
element~\cite{Babich:2009rq,Bali:2009hu,Bali:2009dz,Collins:2010gr,Toussaint:2009pz,Takeda:2010cw}.
The latter may also be determined indirectly from the quark mass
dependence of the nucleon mass via the Feynman-Hellmann
theorem~\cite{Michael:2001bv,Ohki:2008ff,Ohki:2009mt} or SU(3) chiral
perturbation theory~\cite{Young:2009zb}.  Likewise, a complementary
approach for determining the strange electromagnetic form factors
relies on combining lattice data for connected form factors with
chiral perturbation theory using finite range
regularization~\cite{Leinweber:2004tc,Leinweber:2006ug,Wang:1900ta}.
In this work, we present a direct determination of the strange form
factors on relatively large, two-flavor, anisotropic lattices.  Some
preliminary results were presented in~\cite{Babich:2009rq,Babich:2007jg}.

The paper is organized as follows.  In Section~\ref{trace}, we discuss
general considerations for computing the disconnected trace and
describe the particular approach employed in our calculation.  We
discuss our approach for extracting form factors from the
corresponding matrix elements in Section~\ref{corrfunc} and give
further details of the calculation in Section~\ref{details}.  In
Section~\ref{results} we present our results for $G^s_S(Q^2)$,
$G^s_A(Q^2)$, $G^s_E(Q^2)$, and $G^s_M(Q^2)$.  We conclude in
Section~\ref{conclusion} with remarks on the steps we are taking to
improve future determinations of these quantities.

%%%%%%%%%%%%%%%%%%%%%%%%%%%%%%%%%%%%%%%%%%%%%%%%%%%%%%%%%%%%%%%%%%%%%%

\section{Method}

%%%%%%%%%%%%%%%%%%%%

\subsection{\label{trace}Evaluating the trace}

As discussed in Section~\ref{intro}, the evaluation of a disconnected
form factor requires the trace of the quark propagator, times some
combination of Dirac gamma matrices ($\Gamma$) over a time-slice of
the lattice.  The standard method for estimating such a trace relies
on calculating the inverse of the Dirac operator $D$ against an
independent set of random vectors $\eta^{(\alpha)}, \alpha =1,\cdots,
N$ for each spin ($i$), color ($a$), and spatial ($x$) degree of
freedom,
\begin{equation}
\mathrm{Tr}(\Gamma D^{-1})\approx \frac{1}{N} \sum_{\alpha=1}^N
\eta^{^\dagger(\alpha)} \Gamma D^{-1}\eta^{(\alpha)} \, ,\quad \langle\eta_{i,a,x}^*
\eta_{j,b,y}\rangle_\eta = \delta_{xy} \delta_{ab} \delta_{ij}\,.
\label{eq:stochastic}
\end{equation}
Given a finite ensemble of such noise vectors, this procedure
introduces a new source of statistical error, $\pm \sigma/\sqrt{N}$,
measured by the variance $\sigma^2$.  We note
that if we choose an appropriate basis, $\Gamma = (1,\gamma_5,i
\gamma_\mu, \gamma_5 \gamma_\mu ,i\sigma_{\mu \nu})$, for the
Euclidean gamma matrices, $\gamma_5$ invariance implies all the
traces in Eq.~\ref{eq:stochastic} are real. Introducing Gaussian random
vectors, the variance for the real part of $O(\eta)$ for large $N$ is
\begin{eqnarray}
\sigma^2 &=& \frac{1}{4}\left(\< [O^*(\eta) + O(\eta)]^2 \>_\eta - \<
O^*(\eta) + O(\eta) \>^2_\eta\right) \nonumber \\
&=& \frac{1}{2} \sum_{x,y}\mathrm{tr}[D^{\dagger-1}(x,y)  D^{-1}(y,x)]
+ \frac{1}{2} \sum_{x,y}\mathrm{tr} [\Gamma D^{-1}(x,y) \Gamma
D^{-1}(y,x)]\,,
\label{eq:variance}
\end{eqnarray}
where the operators are $O(\eta) =\eta^\dagger \Gamma D^{-1}\eta $ and
$\mathrm{tr}[\cdots]$ now stands for the trace in spin and color
alone.  It is also common in practice to introduce ``unitary'' noise
elements in $U(1)$ or $Z_2$~\cite{Dong:1993pk}, such that $\eta^* \eta
= 1$ which eliminates the diagonal terms corresponding to $x=y$. In
either case, the off-diagonal terms for the first term on the right
hand side of Eq.~\ref{eq:variance} fall off exponentially in the
space-time separation ($|x-y|$) dictated by the lowest Goldstone
pseudoscalar mass in the quark-antiquark channel, giving a divergent
variance in the chiral limit.  Taking the real part reduced this
divergent term by $1/2$ and substituted more rapidly falling
correlators in the second term, except for the pseudoscalar case where
$\Gamma =\gamma_5$.  In this article we employ two modifications to
reduce the variance.  Additional methods will be explored in a
subsequent publication.

First, we choose a random $SU(3)$ gauge transformation $\Omega_x$ as
our unitary stochastic vector, with elements drawn from a uniform
distribution according to the Haar measure.  We also treat the spin
contractions exactly, along the lines of the ``spin explicit method''
described in~\cite{Viehoff:1997wi}.  As a result,
Eq.~(\ref{eq:stochastic}) is replaced by
\begin{equation}
\mathrm{Tr}(\Gamma D^{-1})\approx \frac{1}{N} \sum_{\alpha=1}^N
\mathrm{tr}[\Omega^{^\dagger(\alpha)} \Gamma D^{-1}\Omega^{(\alpha)}] \,,\quad
\< \Omega^{\dagger ab}_x \Omega^{cd}_y \>_\Omega = \delta_{xy}
\delta^{ad} \delta^{bc}/3 
\,,
\end{equation}
and Eq.~(\ref{eq:variance}) by
\begin{equation}
\sigma^2 =  \frac{1}{2} \sum_{x \ne y}\mathrm{tr_c}\left(
\mathrm{tr_s}[D^{\dagger -1}(x,y) \Gamma^\dagger ] \;
\mathrm{tr_s}[\Gamma D^{-1}(y,x)]\right)
 + \frac{1}{2}\sum_{x \ne y} \mathrm{tr_c}\left(
\mathrm{tr_s}[\Gamma D^{-1}(x,y)] \;
\mathrm{tr_s}[\Gamma D^{-1}(y,x)] \right)\,,
\end{equation}
where $\mathrm{tr_c}$ and $\mathrm{tr_s}$ are color and spin traces
respectively.  Note that all 12 spin/color components in the local
term with $x=y$ are removed from the variance.  In addition, the
explicit spin sum results in separating the $SU(3)$ gauge trace from
the Dirac (spin) trace for each propagator. The variance depends on
the individual gamma structure and as a result in general falls off
faster for large $|x-y|$ than the lowest Goldstone mode. To determine
the specific quark/anti-quark channel that contributes, one must
perform a Fierz transformation on each term to put the gamma matrices
in the conventional position for a meson two-point function.  There
will be only one linear combination affected by the Goldstone
mode. The other 15 combinations are determined by massive meson
channels even in the chiral limit.  We leave a more detailed analysis
to a future publication dealing with the light quark sector where this
becomes a more critical issue.

Second, we introduce {\em dilution} to reduce the variance, by
dividing the stochastic source into subsets and estimating the trace
on each subset separately~\cite{Bernardson:1993yg,Foley:2005ac}.  As a
simple illustration consider even/odd dilution, which involves two
subsets,
\begin{equation}
\mathrm{Tr}(\Gamma D^{-1})\approx \frac{1}{N} \sum_{\alpha=1}^N \eta_e^{(\alpha)\dagger}
\Gamma D^{-1}\eta_e^{(\alpha)}+\frac{1}{N} \sum_{\alpha=1}^N \eta_o^{(\alpha)\dagger}
\Gamma D^{-1}\eta_o^{(\alpha)}\, ,
\end{equation}
where $\eta_e^{(\alpha)}$ and $\eta_o^{(\alpha)}$ are non-zero only on
the even and odd sites, respectively, and
$\eta_e^{(\alpha)}+\eta_o^{(\alpha)}=\eta^{(\alpha)}$ gives the
original noise vector.  This may be generalized to a more aggressive
dilution pattern where a larger number of diluted sources is used,
each more sparse. The lighter the quark mass the more aggressively one
should use dilution.  We combine such a scheme with $SU(3)$ unitary
noise and an exact treatment of the spin sum.  Note that had we
instead used dilution over the color index, the resulting variance
would no longer be gauge invariant.

A full calculation involves two sources of statistical error: the
usual gauge noise and the error in the trace. In this investigation
for the strange quark, we largely eliminate the latter error by
calculating a ``nearly exact'' trace on each of four time-slices with
very aggressive dilution. This is accomplished by employing a large
number of sources ($864 \times 12$ for color/spin on a $24^3 \times
64$ lattice) where each source is nonzero on only 16 sites on each of
the four time-slices.  The sites are chosen such that the smallest
spatial separation between them is $6\sqrt{3}a_s$.  With this
aggressive dilution pattern, we find that it is sufficient to use a
single $SU(3)$ source per subset, provided automatically by the random
gauge noise in stochastically independent configurations of our
ensemble.  Any residual contamination, which we observe to be small,
is gauge-variant and averages to zero.  We note that apart from our
use of dilution, this approach corresponds to the ``wall source
without gauge fixing'' method employed in some of the earliest
investigations of disconnected form
factors~\cite{Fukugita:1994fh,Fukugita:1994ba}.

%%%%%%%%%%%%%%%%%%%%

\subsection{\label{corrfunc}Lattice correlation functions}

In Minkowski space, the familiar Dirac and Pauli form factors of the
nucleon, $F_1(Q^2)$ and $F_2(Q^2)$, are implicitly defined by
\begin{equation}
\langle N(p')|J_\mu|N(p)\rangle = \bar u(p')\left[\gamma_\mu F_1(Q^2) +
\frac{i\sigma_{\mu\nu}q^\nu}{2m} F_2(Q^2)\right]u(p) \,.
\label{eq-ff-em}
\end{equation}
Here $|N(p)\rangle$ is a nucleon state with momentum $p$, $u(p)$ is a
nucleon spinor, and we define $Q^2=-q^2$, where $q=p'-p$ is the
4-momentum transfer.  It is often convenient to consider instead the
Sachs electric and magnetic form factors, given by
\begin{equation}
G_E(Q^2)=F_1(Q^2)-\frac{Q^2}{4M^2}F_2(Q^2)
\end{equation}
and
\begin{equation}
G_M(Q^2)=F_1(Q^2)+F_2(Q^2)\,,
\end{equation}
respectively.  The contribution of an individual quark flavor (e.g.,
the strange quark) is defined by replacing the full electromagnetic
current that appears in Eq.~(\ref{eq-ff-em}),
\begin{equation}
J_\mu = \frac{2}{3}\bar u \gamma_\mu u - \frac{1}{3}\bar d \gamma_\mu d
      - \frac{1}{3}\bar s \gamma_\mu s + \ldots\,,
\end{equation}
by $J_\mu^s = \bar s \gamma_\mu s$.  The corresponding Sachs electric
and magnetic form factors are denoted by $G_E^s(Q^2)$ and
$G_M^s(Q^2)$, respectively.

Similarly, the strange quark contribution to the axial form factor
of the nucleon, $G_A^s(Q^2)$, is implicitly given by
\begin{equation}
\langle N(p')|\bar s\gamma_\mu\gamma_5 s|N(p)\rangle = \bar u(p')
\left[\gamma_\mu \gamma_5 G_A^s(Q^2) +
\frac{q_\mu}{2m} \gamma_5 G_P^s(Q^2)\right]u(p) \,.
\end{equation}
In this equation $G_P^s(Q^2)$ denotes the strange quark contribution
to the induced pseudoscalar form factor of the nucleon, which we will
not consider further here.  Finally, we note that the strange scalar form
factor is trivially given by $G_S^s(Q^2)=\langle N(p')|\bar
ss|N(p)\rangle$.  Our main focus will be on the special case of
$Q^2=0$, with the matrix element denoted simply by $\langle N|\bar
ss|N\rangle$.

Our task is to extract these four quantities from appropriately defined
Euclidean correlation functions on the lattice.  We begin by defining the
usual two-point function for the nucleon, with momentum $\vec q$,
\begin{equation}
G^{(2)}(t,t_0;\vec q) = (1+\gamma_4)^{\alpha\beta}
\sum_{\vec x} e^{i\vec q \cdot \vec x}
\langle N^\beta(\vec x,t)\bar{N}^\alpha (\vec 0,t_0)\rangle \,.
\label{eq-g2}
\end{equation}
Here $N^\alpha=\epsilon_{abc} (u_a^T C\gamma_5 d_b) u_c^\alpha$ is the
standard interpolating operator for the proton, with smeared quark
fields, and $(1+\gamma_4)$ projects out the positive-parity state.  In
everything that follows, we always double our statistics by making use
of the invariance of the action under time reversal.  More concretely,
in this case we combine the ``forward-propagating'' correlator
$G^{(2)}(t,t_0;\vec q)$ with the backward-propagating
$G^{(2)}_-(t_0,t;\vec q)$, where the ``$-$'' subscript indicates that
$(1+\gamma_4)$ in Eq.~(\ref{eq-g2}) has been replaced by
$(1-\gamma_4)$.

Next, we define various three-point functions $G_X^{(3)}(t,t',t_0;\vec q)$,
where $X=S,A,E,M$ correspond to the disconnected scalar, axial, electric, and
magnetic form factors, respectively.  These are given by
\begin{equation}
G_S^{(3)}(t,t',t_0;\vec q)= (1+\gamma_4)^{\alpha\beta} \sum_{\vec x,\vec x'} 
e^{i\vec q \cdot \vec x'} \langle N^\beta(\vec x,t) [\bar\psi \psi(\vec x',t')
- \langle\bar\psi \psi(\vec x',t')\rangle] \bar{N}^\alpha (\vec 0,t_0)\rangle 
\label{eq-g3s}
\end{equation}
for the scalar,
\begin{equation}
G_A^{(3)}(t,t',t_0;\vec q) = \frac{1}{3}
\sum_{i=1}^3 \sum_{\vec x,\vec x'} e^{i\vec q \cdot \vec x'}
[-i(1+\gamma_4)\gamma_i\gamma_5]^{\alpha\beta}
\langle N^\beta(\vec x,t) [A_i(\vec x',t')-\langle A_i(\vec x',t')\rangle]
\bar{N}^\alpha (\vec 0,t_0)\rangle
\label{eq-g3a}
\end{equation}
for the axial,
\begin{equation}
G_E^{(3)}(t,t',t_0;\vec q)= (1+\gamma_4)^{\alpha\beta} \sum_{\vec x,\vec x'} 
e^{i\vec q \cdot \vec x'} \langle N^\beta(\vec x,t) [V_4(\vec x',t')
- \langle V_4(\vec x',t')\rangle] \bar{N}^\alpha (\vec 0,t_0)\rangle 
\label{eq-g3e}
\end{equation}
for the electric, and
\begin{equation}
G_M^{(3)}(t,t',t_0;\vec q) = \frac{1}{2n_q}\sum_{\substack{i,j,k\\q_j\neq 0}}
\epsilon_{ijk}\frac{1}{q_j}\sum_{\vec x,\vec x'} e^{i\vec q \cdot \vec x'}
[-i(1+\gamma_4)\gamma_i\gamma_5]^{\alpha\beta}
\langle N^\beta(\vec x,t) [V_k(\vec x',t')-\langle V_k(\vec x',t')\rangle]
\bar{N}^\alpha (\vec 0,t_0)\rangle
\label{eq-g3m}
\end{equation}
for the magnetic.  Here $V_\mu$ and $A_\mu$ denote the vector and
axial currents, and $n_q$ in Eq.~(\ref{eq-g3m}) simply counts the
number of nonzero components of $\vec q$.  For $V_\mu$, we utilize the
conserved current for the Wilson action, given by
\begin{equation}
V_\mu(x+a_\mu\hat\mu/2) = \frac{1}{2}\left[
\bar\psi(x+a_\mu\hat\mu)(\gamma_\mu+1)U_\mu^\dagger(x)\psi(x) +
\bar\psi(x)(\gamma_\mu-1)U_\mu(x)\psi(x+a_\mu\hat\mu)
\right]\,.
\label{eq-vmu}
\end{equation}
The lattice spacing carries a label $\mu$ because we will consider
anisotropic lattices for which the temporal lattice spacing, $a_4
\equiv a_t$, differs from the spatial lattice spacing, $a_1=a_2=a_3
\equiv a_s$.  For convenience, we define $V_\mu(x)$ on a given site of the
lattice by averaging those terms involving the adjacent forward and
backward links, $V_\mu(x)\equiv [V_\mu(x+a_\mu\hat\mu/2) +
  V_\mu(x-a_\mu\hat\mu/2)]/2$; since the spatial index determines the
phase in the Fourier transform, this corresponds to an
$\mathcal{O}(a^2 q^2)$ redefinition of the three-point function.  For
the axial form factor, we will present results computed both from the
analogous point-split current,
\begin{equation}
A_\mu^\mathrm{(p.s.)}(x+a_\mu\hat\mu/2) = \frac{i}{2}\left[
\bar\psi(x+a_\mu\hat\mu)\gamma_\mu\gamma_5 U_\mu^\dagger(x)\psi(x) +
\bar\psi(x)\gamma_\mu\gamma_5 U_\mu(x)\psi(x+a_\mu\hat\mu)
\right]\,,
\label{eq-a-split}
\end{equation}
and the standard local current,
$A_\mu^\mathrm{(local)}(x)=i\bar\psi(x)\gamma_\mu\gamma_5\psi(x)$.

Note that in Eq.~(\ref{eq-g3s})-(\ref{eq-g3m}) we always employ the
vacuum-subtracted value of the current, $[J(\vec x,t)-\langle J(\vec
  x,t)\rangle]$, even though this is only strictly necessary when $J$
is the scalar density, since the expectation value of the others
vanish.  Given finite statistics, however, and an inexact estimate of
the trace, it is possible that using the vacuum-subtracted value gives
reduced statistical errors.  Empirically, we find that for the strange
axial form factor at $Q^2=0$, the two approaches give
indistinguishable results.  At larger momenta, however, uncertainties
for the vacuum-subtracted quantities are noticeably smaller.

To interpret our results, we require an understanding of the
correlation functions given in Eqs.~(\ref{eq-g3s})-(\ref{eq-g3m}) in
terms of the lowest one or two states that dominate at large times.
This is accomplished by performing a spectral decomposition in the
transfer matrix formalism, and for the nucleon two-point function
given in Eq.~(\ref{eq-g2}), we find
\begin{equation}
G^{(2)}(t,t_0;\vec q) = \sum_n 2\left(1+\frac{m_n}{E_n}\right)
Z_n^2(\vec q) e^{-E_n(t-t_0)}\,.
\end{equation}
Here $Z_n(\vec p)$ is defined by $\langle n,\vec p,s|\bar{N}^\alpha
(\vec 0)|0\rangle = Z_n(\vec p)\bar u_s^\alpha(\vec p)$, where
$\bar{N}^\alpha(\vec x)$ is a creation operator for the nucleon,
$|n,\vec p,s\rangle$ is its $n$th eigenstate (with momentum $\vec p$
and polarization $s$), and we have adopted a relativistic
normalization convention for the states: $\langle n', \vec p', s'|n,
\vec p, s\rangle = 2E_n(\vec p)L^3\delta_{s,s'} \delta^{(3)}_{\vec p,
  \vec p'}$.  The momentum dependence in $Z_n(\vec p)$ arises because
we will generally consider extended, rather than point-like,
operators.  We take $n=1$ to label the ground-state proton, and for
later convenience we collect together the coefficients,
\begin{equation}
c_n(\vec q) = 2\left(1+\frac{m_n}{E_n}\right) Z_n^2(\vec q)\,,
\label{eq-cn}
\end{equation}
yielding
\begin{equation}
G^{(2)}(t,t_0;\vec q) = \sum_n c_n(\vec q) e^{-E_n(t-t_0)}\,.
\end{equation}

Similarly, for a generic three-point function involving a current
$J_X(\vec x,t)$,
\begin{equation}
G_X^{(3)}(t,t',t_0;\vec q) = \Gamma_X^{\alpha\beta}\sum_{\vec x, \vec x'}
e^{i\vec q\cdot \vec x'} \langle N^\beta(\vec x,t) J_X(\vec x',t')
\bar{N}^\alpha(\vec 0, t_0)\rangle\,,
\end{equation}
we find
\begin{equation}
G_X^{(3)}(t,t',t_0;\vec q) = \sum_{m,n} j_{nm}(\vec q)
e^{-m_n(t-t')} e^{-E_m(\vec q)(t'-t_0)}\,,
\label{eq-g3j}
\end{equation}
where the coefficients $j_{nm}$ may generally be expressed in terms of
suitable form factors, depending on the current $J_X$ and the
combination of gamma matrices $\Gamma_X$.  The correlation functions
in Eq.~(\ref{eq-g3s})-(\ref{eq-g3m}) have been constructed such that
in each case, the ground-state coefficient $j_{11}$ may be simply
expressed in terms of a single form factor.  In particular, for
$X=S,A,E$, we have
\begin{equation}
j_{11}(\vec q) = 2\left(1+\frac{m_1}{E_1(\vec q)}\right)
Z_1(\vec 0)Z_1(\vec q) G_X^s(Q^2)\,,
\label{eq-j11x}
\end{equation}
where $G_X^s(Q^2)$ is the corresponding strange form factor of the nucleon.
For the magnetic case, the appropriate expression is
\begin{equation}
j_{11}(\vec q) = \frac{2}{E_1(\vec q)} Z_1(\vec 0)Z_1(\vec q) G_M^s(Q^2)\,.
\label{eq-j11m}
\end{equation}

%%%%%%%%%%%%%%%%%%%%%%%%%%%%%%%%%%%%%%%%%%%%%%%%%%%%%%%%%%%%%%%%%%%%%%

\section{\label{details}Details of the calculation}

We work on a $24^3 \times 64$ anisotropic lattice, utilizing an
ensemble of 863 gauge field configurations provided by the Hadron
Spectrum Collaboration~\cite{Bulava:2009jb}.  These were generated
with two degenerate flavors in the sea.  Our gauge and fermion actions
are those defined in Appendix~\ref{aniso}, with coupling
$\beta=6/g^2=5.5$ and bare anisotropy $\xi_0=2.38$.
In~\cite{Bulava:2009jb}, it was found that this value of $\xi_0$
together with $\nu=1$ gives renormalized gauge and fermion anisotropies
that are consistent with $\xi=3$.

The spatial lattice spacing was determined from the Sommer
scale~\cite{Sommer:1993ce} with the parameter $r_0=0.462(11)(4)$~fm,
taken from~\cite{Aubin:2004wf,Bernard:2006wx}, yielding $a_s =
0.108(7)\,\mathrm{fm} = 3a_t$.  For the light quarks, the mass
parameter $m_l^0$ that appears in the action is $m_l^0=-0.4125$.  The
corresponding pion mass is $M_\pi=416(36)$~MeV~\cite{Bulava:2009jb}.

Given the anisotropy, our lattice has a relatively short extent in
time, which has influenced our choice of method.  It has been
conventional in lattice studies to extract the form factors by
considering various ratios of the three- and two-point functions
defined above.  For example, at zero momentum transfer, one finds
\begin{equation}
R_X(t,t',t_0;Q^2=0) \equiv \frac{G_X^{(3)}(t,t',t_0;\vec 0)}
{G^{(2)}(t,t_0;\vec 0)} \rightarrow G_X^s(Q^2=0),
\label{eq-rx}
\end{equation}
for large time separations.  Instead, we have chosen to fit the
three-point function that appears in the numerator of this ratio
directly.  The reasons are two-fold.  First, this allows us to avoid
contamination from backward-propagating states, which are problematic
due to the short temporal extent of our lattice.  At the same time, it
allows us to explicitly take into account the contribution of
(forward-propagating) excited states.

\begin{figure}
\begin{center}
\includegraphics[width=10cm]{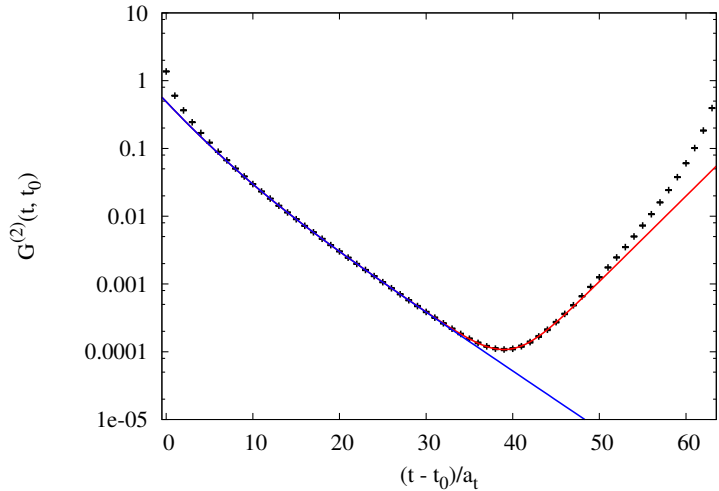}
\caption{\label{fig-prot-corr}Fit of the nucleon correlation function
for $Q^2=0$ to a form that includes two forward-propagating states and
one backward-propagating state (upper curve, red) and the same form
with the coefficient of the backward-propagating set to zero but the other
fit parameters held fixed (lower curve, blue).}
\end{center}
\end{figure}

To see why the direct approach avoids the problem of finite-time
contamination, note that this contamination chiefly affects the nucleon
correlator $G^{(2)}(t,t_0;\vec 0)$ that appears in the denominator of
Eq.~(\ref{eq-rx}), since it involves propagation for a time $(t-t_0)$.
As $(t-t_0)$ exceeds $L_t/2$, the correlator becomes progressively more
contaminated by the negative-parity partner of the nucleon propagating
backward through the lattice.  In Figure~\ref{fig-prot-corr}, we show
a plot of the nucleon correlator, together with a fit to a functional
form that includes two forward-going states and one backward-going
state.  The lower curve shows the effect of dropping the term that
corresponds to the latter; we use this for normalizing some of
our results at zero momentum transfer when we plot them below.

As a result of the contamination in the denominator, the ratio $R_X$
begins to decrease precipitously at large times.  It is important to
note that although $G_X^{(3)}(t,t',t_0;\vec q)$ in the numerator also
involves a nucleon propagating for time $(t-t_0)$, the contamination
there is a concern only insofar as it increases the statistical error
by washing out the correlation we are attempting to measure.  It
remains unbiased since the current is inserted at $t'$, while the
negative-parity partner propagates across the opposite side of the
lattice, from $t$ to $t_0$, and can be expected to correlate little
with the disconnected insertion.

As described in the previous section, the correlation functions given in
Eqs.~(\ref{eq-g3s})-(\ref{eq-g3e}) have been defined such that a single
form factor enters the coefficient $j_{11}$ for each case, according
to Eq.~(\ref{eq-j11x}).  In terms of the coefficients $c_n$ extracted
from the two-point function $G^{(2)}(t,t_0;\vec q)$, this becomes
\begin{equation}
j_{11}(\vec q)=G_X^s(Q^2) \sqrt{\frac{1}{2}\left(1+\frac{m_1}{E_1(\vec q)}
\right) c_1(\vec 0) c_1(\vec q)}
\label{eq-j11x-c}
\end{equation}
for $X=S,E,A$, where $G_X^s(Q^2)$ is the corresponding strange form
factor of the nucleon.
The corresponding expression for $G_M^{(3)}(t,t',t_0;\vec q)$  is
\begin{equation}
j_{11}(\vec q)=\frac{G_M^s(Q^2)}{E_1(\vec q)+m_1}
\sqrt{\frac{1}{2}\left(1+\frac{m_1}{E_1(\vec q)} \right)
c_1(\vec 0) c_1(\vec q)} \,.
\label{eq-j11m-c}
\end{equation}

Our general strategy will be to fit the correlation functions
$G_X^{(3)}(t,t',t_0;\vec q)$ to Eq.~(\ref{eq-g3j}), taking into
account both the ground state nucleon and a single excited state.  We
may then extract the nucleon form factors from $j_{11}$ with input
from the two-point function.  In principle, one could also obtain form
factors of the first excited state from $j_{22}$, as well as
transition form factors from $j_{12}$ and $j_{21}$.  In practice,
however, we expect these to absorb the contributions of still higher
states and trust only the ground state form factors to be reliable.

%%%%%%%%%%%%%%%%%%%%%%%%%%%%%%%%%%%%%%%%%%%%%%%%%%%%%%%%%%%%%%%%%%%%%%

\section{\label{results}Results and Discussion}

%%%%%%%%%%%%%%%%%%%%

\subsection{\label{scalar}Strange scalar form factor and $f_{Ts}$}

\begin{figure}
\begin{center}
\includegraphics[width=10cm]{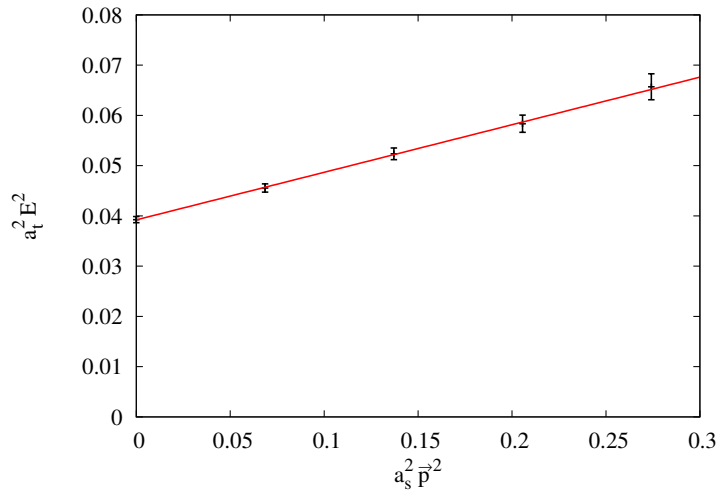}
\caption{\label{fig-dispersion}Nucleon energy-squared (in lattice
  units) as a function of momentum, together with a fit to the
  continuum dispersion relation.}
\end{center}
\end{figure}

For all of the results presented in this section, the nucleon
two-point function was fit in the range $10\le t/a_t\le 45$, yielding
$a_t M_N=0.198(2)$ for the ground-state nucleon mass.  In
Figure~\ref{fig-dispersion}, we plot the nucleon energy (squared) as a
function of momentum for the five smallest values of $|\vec{p}|^2$
available on our lattice, along with a fit to the continuum dispersion
relation $(a_t E)^2 = (a_s|\vec{p}|)^2/\xi^2 + (a_t m)^2$.  The
parameter $\xi$, given by the inverse square root of the slope,
provides a measure of the effective fermion anisotropy $a_s/a_t$.  We
find $\xi=3.25(11)$ from the fit, which may be compared to the values
2.979(28) and 3.045(35) obtained from the pion and rho dispersion
relations, respectively, in~\cite{Bulava:2009jb}.  The intercept $(a_t
m)^2$ in Figure~\ref{fig-dispersion} is largely constrained by the point
at $|\vec{p}|^2=0$ and thus yields an identical value (and error) for
the nucleon mass.

\begin{figure}
\begin{center}
\includegraphics[width=10cm]{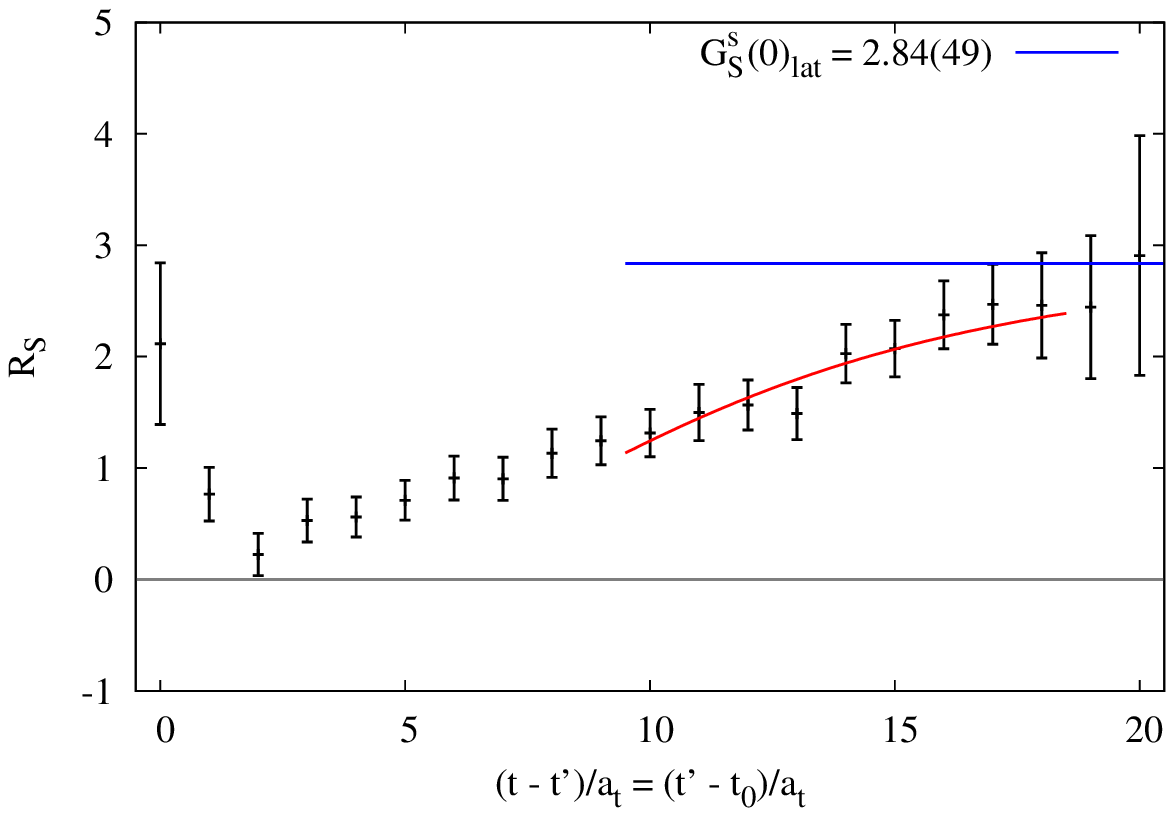}
\caption{\label{fig-scalar-q0}Subset of results for the scalar form factor
at $Q^2=0$, where the current insertion is placed symmetrically between
source and sink.  The lower curve (red) shows a corresponding cross-section
of the fit.  The horizontal line (blue) indicates the resulting value of
$G_S^s(Q^2=0)_\mathrm{lat}=\langle N|\bar ss|N\rangle_0$ for the ground-state
nucleon.}
\end{center}
\end{figure}

\begin{figure}
\begin{center}
\includegraphics[width=10cm]{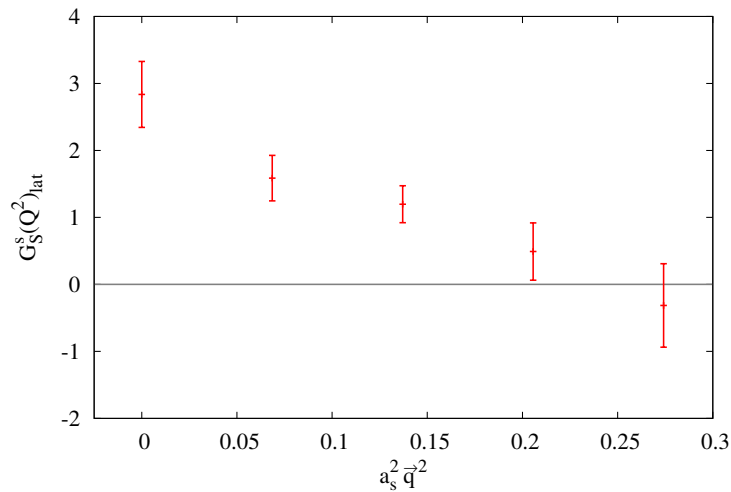}
\caption{\label{fig-scalar}Strange scalar form factor as a function of
  momentum.}
\end{center}
\end{figure}

In order to extract the form factors, the various three-point
functions were fit to Eq.~(\ref{eq-g3j}), taking into account the two
lowest-lying states, with the separations $(t-t')/a_t$ and
$(t'-t_0)/a_t$ varying independently in the range $[10,18]$.  It
follows that a total of 81 data points are included in the fit.  A
one-dimensional subset of these points for the scalar form factor at
$Q^2=0$ is shown in Figure~\ref{fig-scalar-q0}.  For the purpose of
plotting, we have normalized our results by a fit to the two-point
function.  With this normalization, dominance of the ground state
should manifest as a plateau at large times.  We find
$G_S^s(0)_\mathrm{lat} = 2.84(49)$ for the form factor at zero
momentum transfer, where the statistical error has been determined via
a single-elimination jackknife applied to the full fitting procedure.
In Figure~\ref{fig-scalar}, we show the momentum dependence of the
strange scalar form factor.

We discuss the systematic uncertainties affecting these results below,
including the delicate problem of relating the bare matrix element to
the continuum.  One practical consideration is the choice of fitting
windows used in the fits of the two- and three-point functions.
In order to extract the form factors from
$G_X^{(3)}(t,t',t_0;\vec q)$, we must first determine the coefficients
$c_n(\vec q)$ and masses/energies $E_n(\vec q)$ from a fit to
$G^{(2)}(t,t_0;\vec q)$.  Since we have access to a total of $863
\times 64 = 55,232$ nucleon correlators, these tend to be very
well-determined, as illustrated by
Figure~\ref{fig-prot-corr}.  The coefficients $c_n$ are somewhat
sensitive to the choice of fitting window, however, and since they
multiply the form factor in Eq.~(\ref{eq-j11x-c}), this translates
into a direct systematic error on the form factor, estimated to be
about ten percent.

\begin{figure}
\begin{center}
\includegraphics[width=10cm]{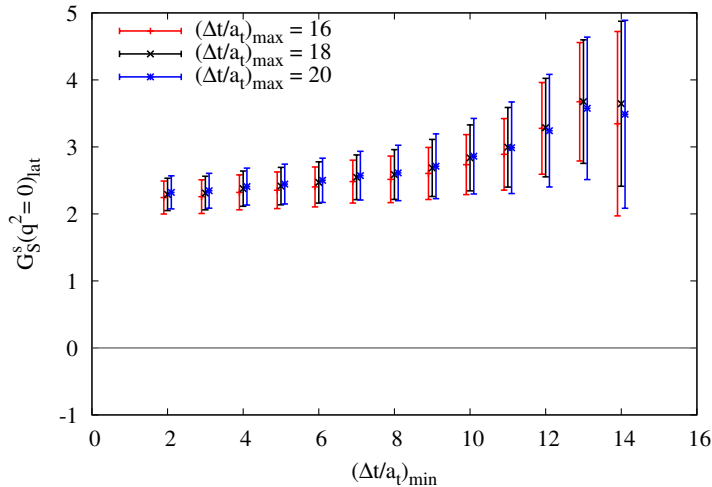}
\caption{\label{fig-scalar-scan}Dependence of the extracted value of
$G_S^s(Q^2=0)_\mathrm{lat}=\langle N|\bar ss|N\rangle_0$ on the range
of time separations included in the fit.}
\end{center}
\end{figure}

In contrast, we find that our results are relatively insensitive to
the choice of window used in the fit of the three-point function.
This is illustrated in Figure~\ref{fig-scalar-scan}, where we plot the
extracted value of $G_S^s(Q^2=0)_\mathrm{lat}=\langle N|\bar
ss|N\rangle_0$ as a function of the smallest time separation included
in the fit, for three different values of the maximum time separation.
(See also the analogous plot for $G_A^s(Q^2=0)$,
Figure~\ref{fig-axial-scan} below.)  We observe a stable plateau that
extends to very early time separations but have nevertheless chosen a
conservative lower bound, $(t-t') \ge 10a_t$ and $(t'-t_0)\ge 10a_t$,
effectively eliminating systematics due to excited-state contamination
of the three-point function, at the expense of increased statistical
errors.

\begin{figure}
\begin{center}
\includegraphics[width=0.6\textwidth]{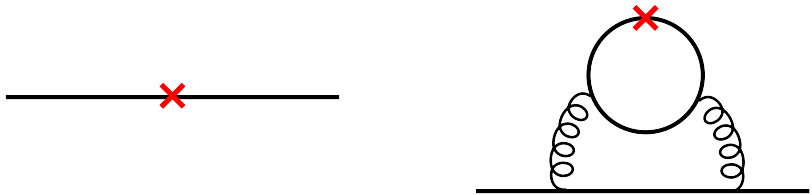}
\caption{\label{fig-mixing}The left diagram contributes to the
  renormalization of both the flavor singlet and flavor non-singlet
  mass operator, while the right appears only for the flavor
  singlet case.}
\end{center}
\end{figure}

Thus far, we have computed only the unrenormalized matrix element
$\langle N|\bar ss|N\rangle_0$ on the lattice.  Naively we can
multiply by the subtracted bare strange quark mass $\widetilde m^0_s =
m^0_s - m_\mathrm{crit}$ (determined in Appendix~\ref{quarkmass}), to find
$(\sigma_s)_\mathrm{lat}=504(91)(30)$~MeV, which in the continuum
corresponds to the renormalization group invariant quantity
$\sigma_s=m_s\langle N|\bar ss|N\rangle$.  Here the second error
reflects the uncertainty in the lattice scale, the first is
statistical, and no other systematics have been taken into account. If
we then divide by our measured value of the nucleon mass, $a_t
M_N=0.198(2)$, the lattice spacing dependence drops out, yielding
$(f_{Ts})_\mathrm{lat} = \widetilde m^0_s \langle N|\bar
ss|N\rangle_0/M_N = 0.46(9)$.  This large value is in apparent
disagreement with recent lattice determinations using staggered and
chiral
fermions~\cite{Toussaint:2009pz,Takeda:2010cw,Ohki:2008ff,Ohki:2009mt}.
The source of this discrepancy, as first pointed out
in~\cite{Michael:2001bv}, is the explicit breaking of chiral symmetry
in the Wilson action, which allows for mixing between singlet and
non-singlet matrix elements even after tuning the quark masses
$\widetilde m^0_i = m^0_i - m_\mathrm{crit}$ to zero.  As a consequence, at
finite lattice spacing the strange scalar matrix element receives
contributions from both connected and disconnected diagrams involving
the {\em light} quarks, such as those illustrated in
Figure~\ref{fig-mixing}.

As described in~\cite{Rakow:2004vj,Gockeler:2004rp} in the context of
quark mass renormalization, a natural approach for treating this
problem is to separately consider the renormalization of flavor
singlet and non-singlet contributions.  In
Appendix~\ref{mixing}, we employ this approach with the aid of the
lattice Feynman-Hellmann theorem to rederive a result recently quoted
in~\cite{Takeda:2010cw} for the renormalized matrix element,
\begin{equation}
\langle N|\bar ss|N\rangle =  \frac{1}{3} \left[(Z_0 + 2 Z_8)\langle N|\bar
ss |N\rangle_0 + (Z_0 -Z_8) \langle N| \bar u u + \bar d d|N\rangle_0 \right]
+ c \langle N| \mathrm{Tr}[F^2] |N\rangle_0\,.
\label{eq:CondRen}
\end{equation}
Here $Z_0$ and $Z_8$ are the flavor singlet and non-singlet
renormalization constants for the scalar density, $\mathrm{Tr}[F^2]$
is the gauge kinetic term, and $c$ is a constant.  The discussion in
Appendix~\ref{mixing} closely parallels the analysis of Bhattacharya,
Gupta, Lee, Sharpe, and Wu~\cite{Bhattacharya:2005rb}, who consider in
detail operator mixing for $N_f = 2+1$ clover-improved Wilson fermions
with $m_d = m_u < m_s$, including all terms to $O(a)$ and $O(a
m_q)$. This is a generalization of the classic on-shell $O(a)$
improvement scheme of the ALPHA collaboration~\cite{Jansen:1995ck,
  Luscher:1996vw}.

A self-consistent application of this approach demands an
$O(a)$-improved action with $2+1$ dynamical flavors in the sea; such a
calculation is under way (see conclusion) but beyond the scope of this
paper.  Nonetheless the discussion in Appendix~\ref{mixing} is
intended to clarify the source of the mixing problem.  In accordance
with~\cite{Rakow:2004vj,Gockeler:2004rp,Bhattacharya:2005rb}, it
demonstrates that the singlet ($Z^m_0$) and non-singlet ($Z^m_8$) mass
renormalization constants separately obey the reciprocal relations
$Z^m_0 = 1/Z_0$ and $Z^m_8 = 1/Z_8$ at zero quark mass and indicates
why one expects the gluonic mixing (parameterized by $c$) to be small.
In the approximation where the gluonic mixing is neglected ($c=0$),
correcting the dimensionless ratio $f_{Ts} = m_s \langle N| \bar s s|
N \rangle/M_N$ only requires computation of the ratio $Z_8/Z_0$.
Consequently, we believe the value for the renormalization of the
condensates can in principle be estimated following the prescription
outlined in~\cite{Rakow:2004vj,Gockeler:2004rp}; by varying the
valence and sea quark mass separately one can separate out the singlet
and non-singlet contributions.  Alternatively, one could determine
this ratio by evaluating singlet and non-singlet matrix elements
directly.  Details of how best to compute the corrections are left for
a future work.  Suffice it to say that since the corrections due to
mixing are large and negative (i.e., $Z_8/Z_0 > 1$), we cannot rule
out the possibility that the renormalized quantity $\langle N|\bar
ss|N\rangle$ is consistent with zero within errors for the present
calculation.

\begin{figure}
\begin{center}
\includegraphics[width=10cm]{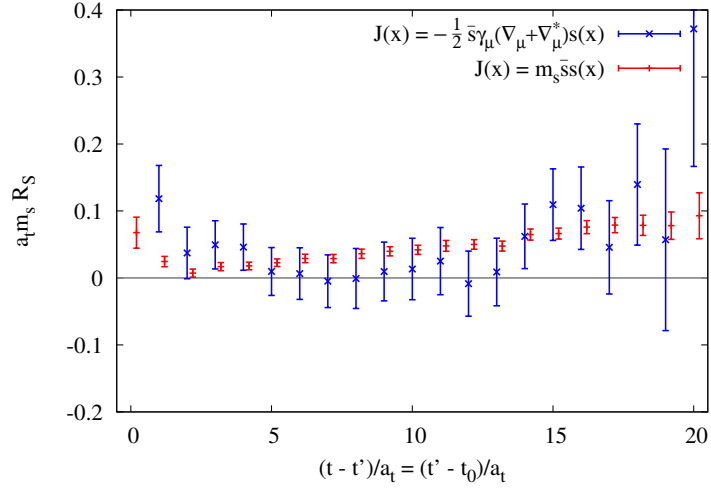}
\caption{\label{fig-kinetic-q0}Determination of $\sigma_s=m_s\langle
  N|\bar ss|N\rangle$ (in lattice units) from the discretized ``kinetic term,'' as
  compared to same quantity evaluated by direct insertion of the
  scalar density multiplied by the subtracted bare quark mass.}
\end{center}
\end{figure}

\begin{figure}
\begin{center}
\includegraphics[width=10cm]{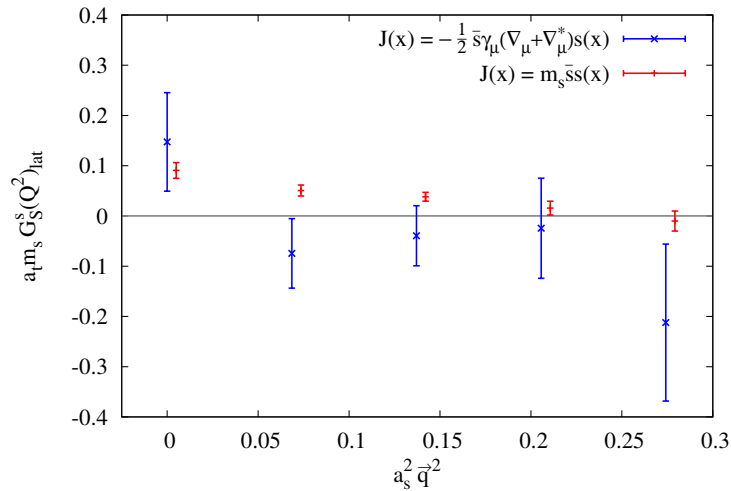}
\caption{\label{fig-kinetic}Momentum dependence of $a_t m_s G_S^s(Q^2)$, as
  determined directly and from the kinetic term.}
\end{center}
\end{figure}

We now consider an alternative method for determining
$\sigma_s=m_s\langle N|\bar ss|N\rangle$ by invoking the continuum
equations of motion to replace $m_s\bar ss$ by the quark ``kinetic
term,'' $\bar s\gamma_\mu D_\mu s$.  On the lattice, the covariant derivative
goes over to $D_\mu=(\nabla_\mu+\nabla_\mu^*)/2$, defined in terms of the
covariant finite difference operators $\nabla_\mu$ and $\nabla_\mu^*$
given in Appendix~\ref{aniso}.  Evaluating the matrix element of this
operator gives us an alternative determination of $\sigma_s$, with lattice
artifacts that are at least different and potentially less severe than
those affecting the direct approach.  In particular, this alternative obviates
the need to separately consider quark mass and operator subtractions.
Before turning to our results, we note that by splitting the lattice
Wilson Dirac operator into three pieces,
\begin{equation}
D = \gamma_\mu D_\mu[U] + m_0^i + W[U]\,
\end{equation}
corresponding to the kinetic term, the bare mass term, and the Wilson
term, respectively, and by invoking the exact lattice equation of
motion, we may write this matrix element in two equivalent ways:
\begin{equation}
- \langle N |\bar s\gamma_\mu D_\mu s|N\rangle = \langle N|(m^0_s \bar
s s + \bar s W s) |N \rangle \,.
\end{equation} 
From the expression on the right, we see that we have in effect
subtracted the major shift due to $m_\mathrm{crit}$.  Indeed an alternative
definition of the critical mass follows from imposing the condition,
\begin{equation}
 \langle N | (m^0_s  \bar s s + \bar s W s) |N \rangle  = (m^0_s - \hat m_\mathrm{crit}) \langle N |  \bar s s|N \rangle
\end{equation} 
where $\hat m_\mathrm{crit} \equiv -\langle N | \bar s W s |N \rangle /\langle
N | \bar s s |N \rangle $.  With this definition, at $m^0_s =\hat
m_\mathrm{crit}$ the flavor singlet term in Figure~\ref{fig-mixing} is set to
zero, suggesting that this scheme may suffer from smaller operator mixing
than the direct approach.

A practical question is how the statistical uncertainties in the two
approaches compare.  In Figure~\ref{fig-kinetic-q0}, we show our
results for $\sigma_s$ determined from the matrix element of the
kinetic term.  For comparison, we also include the data for $\langle
N|\bar ss|N\rangle_0$ shown previously in Figure~\ref{fig-scalar-q0},
but now rescaled by the bare subtracted quark mass (using the standard
definition computed in Appendix~\ref{quarkmass}).  If not for lattice
artifacts, these two sets of results would correspond to the same
continuum quantity.  We find that the determination from the kinetic
term does in fact suffer from much larger statistical errors, perhaps
limiting the usefulness of the approach.  A final judgement should
await comparison of properly subtracted and renormalized results; such
an investigation is underway.  For completeness,
Figure~\ref{fig-kinetic} shows results for the scalar form factor as a
function of momentum using the two approaches.  The points with
smaller error bars correspond to the data of Figure~\ref{fig-scalar},
rescaled by the bare subtracted quark mass.

%%%%%%%%%%%%%%%%%%%%

\subsection{Strange axial form factor and $\Delta s$}

\begin{figure}
\begin{center}
\includegraphics[width=10cm]{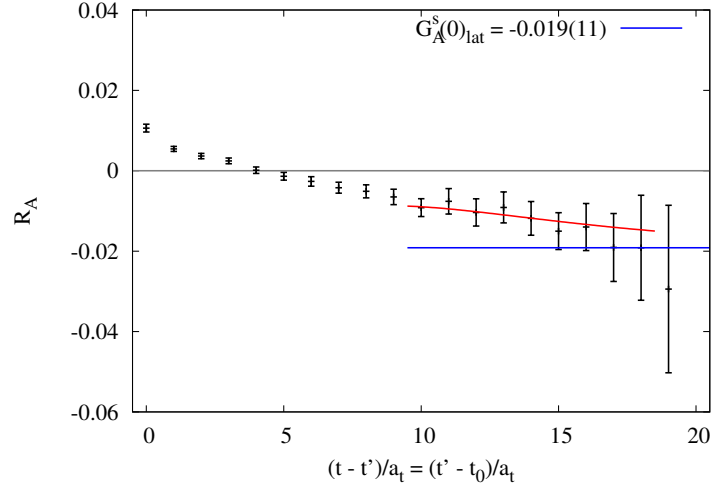}
\caption{\label{fig-axial-q0}Subset of results for the axial form factor
at $Q^2=0$, where the current insertion is placed symmetrically between
source and sink.  The lower curve (red) shows a corresponding cross-section
of the fit.  The horizontal line (blue) indicates the resulting value of
$G_A^s(Q^2=0)_\mathrm{lat}=(\Delta s)_\mathrm{lat}$ for the ground-state
nucleon.}
\end{center}
\end{figure}

\begin{figure}
\begin{center}
\includegraphics[width=10cm]{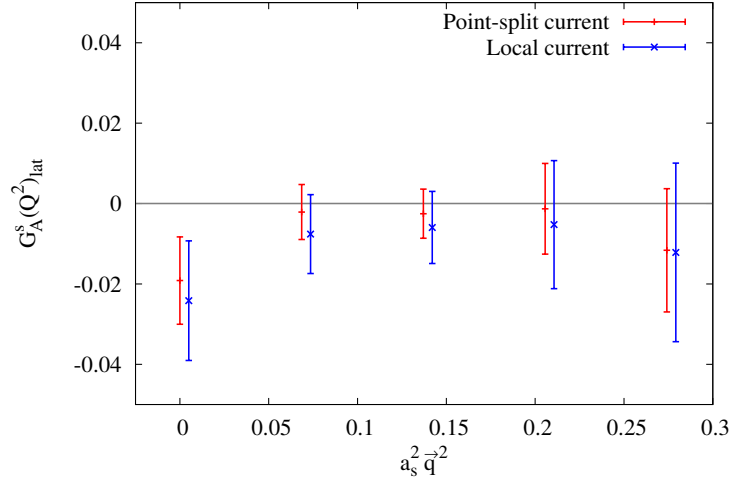}
\caption{\label{fig-axial}Strange axial form factor as a function of
  momentum.}
\end{center}
\end{figure}

Results for the strange axial form factor are shown in
Figure~\ref{fig-axial-q0}, here computed using the point-split current
of Eq.~(\ref{eq-a-split}).  As was the case for $G_S^s(Q^2)$, we note
that our result $(\Delta s)_\mathrm{lat} = G_A^s(0)_\mathrm{lat} =
-0.019(11)$ has not been renormalized and so may not be compared
directly to experimental results.\footnote{We also note that the
  preliminary results for the bare quantity $G_A^s(0)_\mathrm{lat}$
  reported in~\cite{Babich:2009rq} were computed using a point-split
  current involving gauge links rescaled by the bare anisotropy,
  $\xi_0=2.38$.  Here we adopt a more conventional normalization for
  the current, for which $Z_A\sim 1$.}  Despite the large errors, the
data in Figure~\ref{fig-axial-q0} seem to strongly favor a negative
value for $\Delta s$, an observation that is in itself of
phenomenological interest, given the present uncertainties in
experimental determinations and the continued disagreement among some
model calculations over the sign.  In Figure~\ref{fig-axial}, we show
the momentum dependence of $G_A^s(Q^2)$.  Because the determination of
renormalization constants for our anisotropic lattice action is still
pending, we present results for both the point-split and local axial
currents.

\begin{figure}
\begin{center}
\includegraphics[width=10cm]{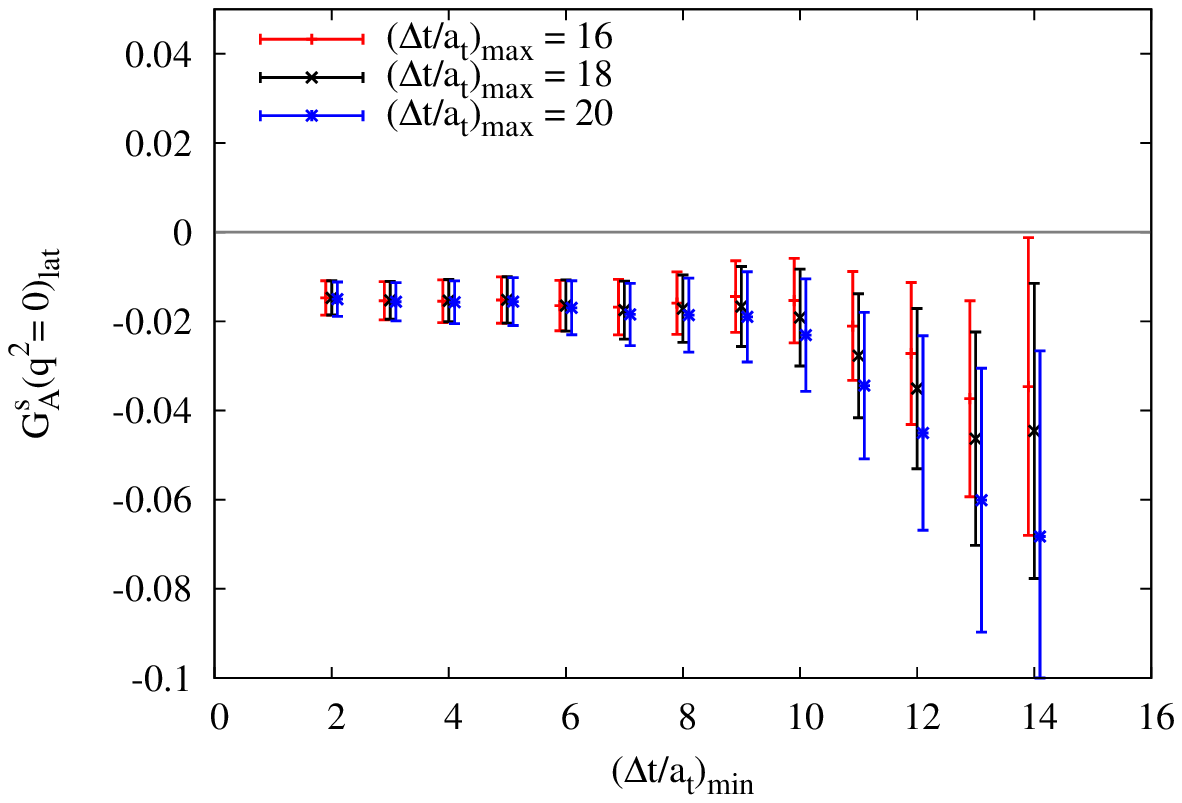}
\caption{\label{fig-axial-scan}Dependence of the extracted value of
  $G_A^s(Q^2=0)_\mathrm{lat}$ on the range of time separations included in the
  fit.}
\end{center}
\end{figure}

Finally, Figure~\ref{fig-axial-scan} shows how the extracted value of
$G_A^s(Q^2=0)$ would vary as a function of the range of time separations
included in the fit.  As described earlier, in computing all our
results we have chosen a conservative range with the time separations
$(t-t')/a_t$ and $(t'-t_0)/a_t$ varying independently in the interval
$[10,18]$.  The corresponding point in the figure is labeled by
$(\Delta t/a_t)_\mathrm{min}=10$ and $(\Delta t/a_t)_\mathrm{max}=18$.

%%%%%%%%%%%%%%%%%%%%

\subsection{Strange electric and magnetic form factors}

\begin{figure}
\begin{center}
\includegraphics[width=10cm]{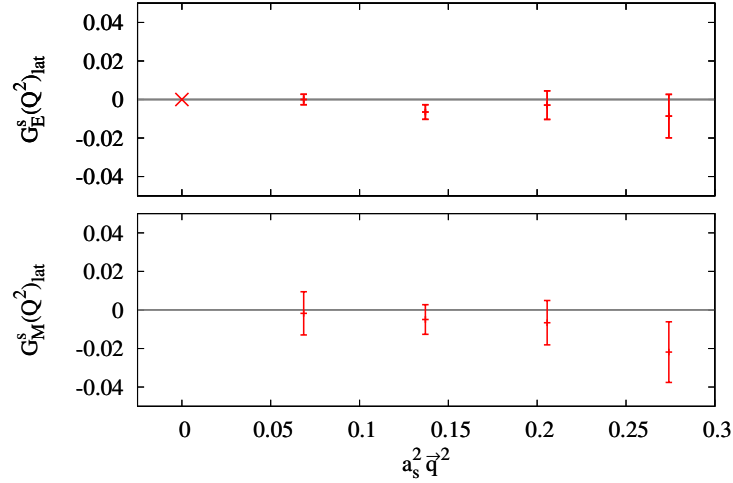}
\caption{\label{fig-em}Strange electric and magnetic form factors as a function
  of momentum.}
\end{center}
\end{figure}

In Figure~\ref{fig-em}, we present our results for the strange quark
contribution to the nucleon's electric and magnetic form factors, as a
function of momentum.  We have used the vector current defined in
Eq.~\ref{eq-vmu}, which is conserved for the Wilson action and
therefore does not get renormalized.  Note that since the strange
quark does not contribute to the electric charge of the nucleon,
$G_E^s(Q^2=0)$ must vanish.  This provides an additional check of our
method, and we find $G_E^s(Q^2=0)=-0.0016(20)$, consistent with zero
as expected.  More generally, all of our results for $G_E^s(Q^2)$ and
$G_M^s(Q^2)$ appear to be roughly consistent with zero, implying that
these quantities are rather small for $Q^2>0.1\ \mathrm{GeV}^2$.
Strictly speaking, we cannot set proper limits without extrapolating
our results to the continuum and to the physical value of the light
quark mass, but it is notable that the statistical errors are as much
as an order of magnitude smaller than the corresponding experimental
uncertainties (cf.~\cite{Pate:2008va,Baunack:2009gy}).  This suggests
that measuring a nonzero value for $G_{E,M}^s(Q^2)$ in electron
scattering experiments may be a challenging task indeed.

\begin{table}
\begin{center}
\begin{tabular}{@{\extracolsep{1em}}cl|lllll}
\hline\hline
$(L_s/2\pi)^2|\vec q|^2$ & $Q^2\ [\mathrm{GeV}^2]$
& $G_S^s(Q^2)_\mathrm{lat}$
& $G_A^s(Q^2)_\mathrm{lat}^\mathrm{(p.s.)}$
& $G_A^s(Q^2)_\mathrm{lat}^\mathrm{(local)}$
& $G_E^s(Q^2)_\mathrm{lat}$
& $G_M^s(Q^2)_\mathrm{lat}$ \\
\hline
0 & 0        & \phantom{-}2.84(49) & -0.019(11) & -0.024(15) & \ \ \ \ \ ---       & \ \ \ \ \ --- \\
1 & 0.22(3)  & \phantom{-}1.59(34) & -0.002(7)  & -0.008(10) & \phantom{-}0.000(3) & -0.002(11) \\
2 & 0.43(5)  & \phantom{-}1.20(28) & -0.003(6)  & -0.006(9)  & -0.007(4)           & -0.005(8) \\
3 & 0.62(8)  & \phantom{-}0.49(43) & -0.001(11) & -0.005(16) & -0.003(7)           & -0.007(12) \\
4 & 0.81(10) &           -0.31(62) & -0.012(15) & -0.012(22) & -0.009(11)          & -0.022(16) \\
\hline\hline
\end{tabular}
\end{center}
\caption{\label{tab-formfac}Summary of results for strange form factors of
the nucleon.}
\end{table}

In Table~\ref{tab-formfac}, we summarize our results for the strange
form factors of the nucleon, with the momentum transfer $Q^2$ given by
\begin{equation}
%Q^2 = |\vec q|^2 - \left(M_N-\sqrt{|\vec q|^2+M_N^2}\right)^2\,,
Q^2 = 2 M_N \left( \sqrt{|\vec q|^2+M_N^2} - M_N \right)\,,
\end{equation}
where $M_N$ is our lattice determination of the nucleon mass.  The quoted
errors for $Q^2$ reflect the uncertainties in $M_N$ and the lattice
scale.  We again emphasize that the results in Table~\ref{tab-formfac} were
determined with $m_{u,d}$ unphysically heavy, corresponding to a pion
mass of about 400~MeV, and that the tabulated values for $G_S^s(Q^2)$
and $G_A^s(Q^2)$ are unrenormalized.

%%%%%%%%%%%%%%%%%%%%%%%%%%%%%%%%%%%%%%%%%%%%%%%%%%%%%%%%%%%%%%%%%%%%%%

\section{Conclusion}
\label{conclusion}

In this work, we have described our first effort to compute
disconnected contributions to nucleon form factors, focusing on the
strange quark.  Employing the Wilson gauge and fermion actions on an
anisotropic lattice, we computed a large number of nucleon correlators
and accurate unbiased estimates for the disconnected currents on each
gauge configuration.  We nevertheless found results for the
electromagnetic form factors that are consistent with zero and a
result for $\Delta s$ that is only marginally distinct from zero,
suggesting that the physical values of these quantities are rather
small.  Such null results may be interpreted as limits --- with the
aforementioned caveats concerning systematics --- and should also be
useful for setting bounds on the disconnected contributions that are
generally neglected in lattice determinations of nucleon form factors
(or explicitly canceled by taking isovector combinations).  To
complete this program, it will of course be necessary to include
disconnected contributions from light quarks as well.

In the future, we plan to build on the present investigation by
introducing several improvements.  First, we are making use of
multiple ensembles of anisotropic lattices with 2+1 flavors in the
sea~\cite{Lin:2008pr}, which will allow the strange quark to be
treated fully self-consistently.  These were generated with a Wilson
fermion action that is stout-smeared~\cite{Morningstar:2003gk} and
$O(a)$-improved~\cite{Sheikholeslami:1985ij}, both features that may
be expected to improve the chiral properties of the
action~\cite{Capitani:2006ni} and thereby reduce the effect of flavor
mixing discussed in Section~\ref{scalar}.  Indeed, the fact that their
action is clover-improved may explain why the authors
of~\cite{Collins:2010gr} found a value for $\langle N|\bar
ss|N\rangle_0$ that is significantly smaller than ours (but still
larger than determinations employing chiral or staggered fermions);
one must also take multiplicative renormalization factors into account
when comparing bare values obtained with different actions, but such
factors are not expected to differ enough from unity to account for
the discrepancy.  These new ensembles also have a much longer extent
in time (with volumes of $24^3\times 128$ and larger), which will
suppress contaminations from backward-propagating states and allow us
to obtain a signal over a larger range of time separations, thus
reducing statistical errors.

Second, we are leveraging a powerful new adaptive multigrid (MG)
algorithm for inverting the Wilson-clover Dirac operator that is
allowing us to compute the disconnected diagrams for both strange and
light quarks at very little additional
cost~\cite{Babich:2010qb,Osborn:2010mb}.  We are also taking advantage
of clusters accelerated by graphics processing units (GPUs) using the
QUDA library~\cite{Clark:2009wm,Babich:2010mu}, which provides another
substantial speedup.  Work is underway to develop an MG implementation
suitable for GPUs, in lieu of the Krylov solvers currently implemented
in QUDA.  We estimate that by combining these two improvements, we may
be able to reduce the cost per Dirac inversion at light quark masses
by up to two orders of magnitude as compared to standard solvers on
traditional architectures.  Finally, we are exploring additional
methods for reducing the variance in estimates of the trace of
disconnected currents, such as the multigrid subtraction method
described in~\cite{Babich:2007jg}.  The net result of these
improvements will be a significant reduction in both statistical and
systematic errors.  At the same time, the scheme outlined in
Appendix~\ref{mixing} should allow us to correct for operator mixing
in the determination of the strange scalar matrix element, yielding a
reliable value and further elucidating the connection between results
obtained with chiral and Wilson-like fermions.

%%%%%%%%%%%%%%%%%%%%%%%%%%%%%%%%%%%%%%%%%%%%%%%%%%%%%%%%%%%%%%%%%%%%%%

\begin{acknowledgments}
We wish to acknowledge useful discussions with Joel Giedt and Stephen Sharpe.
This work was supported in part by U.S.\ DOE grants DE-FG02-91ER40676
and DE-FC02-06ER41440; NSF grants DGE-0221680, PHY-0427646, and
PHY-0835713; and by the NSF through TeraGrid resources provided by the
Texas Advanced Computing Center~\cite{catlett2007tao}.  Computations
were also carried out on facilities of the USQCD Collaboration, which
are funded by the Office of Science of the U.S.\ Department of Energy,
as well as on the Scientific Computing Facilities of Boston University.
\end{acknowledgments}

%%%%%%%%%%%%%%%%%%%%%%%%%%%%%%%%%%%%%%%%%%%%%%%%%%%%%%%%%%%%%%%%%%%%%%

\appendix

\section{\label{aniso}The Wilson action for an anisotropic lattice}

In our calculation, we take the temporal lattice spacing $a_t$ to be
finer than that in the three spatial directions, which share a common
value $a_s$.  In the interacting theory, the anisotropy $\xi \equiv
a_s/a_t$ renormalizes away from the bare value that appears in the
action, which we denote by $\xi_0$.  Furthermore, the anisotropies
appearing in the gauge and fermion actions may in principle
renormalize differently.  We follow~\cite{Bulava:2009jb} in denoting
the bare gauge anisotropy by $\xi_0$ while introducing a new parameter
$\nu$ such that the bare fermion anisotropy is given by $\xi_0/\nu$.
We assume that the renormalization of the latter quantity is
independent of quark mass, as found empirically
in~\cite{Bulava:2009jb,Edwards:2008ja}.

With these definitions, the Wilson gauge action on an anisotropic lattice
is given by~\cite{Klassen:1998ua}
\begin{equation}
S_g =
\frac{6}{\xi_0 g^2}\sum_x\sum_{\mu=1}^3\left[
\sum_{\mu<\nu<4}\left(1-\frac{1}{3}\mathrm{Re}\,U_{\mu\nu}(x)\right)
+\xi_0^2 \left(1-\frac{1}{3}\mathrm{Re}\,U_{\mu 4}(x)\right)
\right]\,,
\end{equation}
in terms of the plaquette
$U_{\mu\nu}(x) = \mathrm{Tr}\left[U_\mu(x)U_\nu(x+a_\mu\hat\mu)
U_\mu^\dagger(x+a_\nu\hat\nu)U_\nu^\dagger(x)\right]$,
where $\mu=4$ corresponds to the ``time'' direction.  The Wilson fermion
action, in turn, is given by~\cite{Klassen:1998fh}
\begin{eqnarray}
S_W=a_s^3\sum_x\bar\psi(x)\Bigg[a_t m_q^0 &+&
 \frac{\nu}{\xi_0}a_s\sum_{i=1}^3
\left( \frac{1}{2}\gamma_i(\nabla_i+\nabla_i^*)
-\frac{a_s}{2}\nabla_i^*\nabla_i \right)
+ a_t\left(\frac{1}{2}\gamma_4(\nabla_4+\nabla_4^*)
-\frac{a_t}{2}\nabla_4^*\nabla_4\right)
\Bigg]\psi(x)\,,
\label{eq-sw-aniso}
\end{eqnarray}
where we have defined the covariant difference operators
$\nabla_\mu \psi(x) = [U_\mu(x)\psi(x+a_\mu\hat\mu) - \psi(x)]/a_\mu$
and
$\nabla^*_\mu \psi(x) = [\psi(x)
-U^\dagger_\mu(x-a_\mu\hat\mu)\psi(x-a_\mu\hat\mu)]/a_\mu$.
Note that $a_t$ occurs in Eq.~(\ref{eq-sw-aniso}) only to cancel where
it appears in the definition of $\nabla_4$, except in the
dimensionless mass parameter $(a_t m_q^0)$.  We can write the fermion
action in a more familiar and explicit form by defining rescaled
links,
\begin{equation}
\tilde U_\mu(x) = \left\{ \begin{array}{rl}
\frac{\nu}{\xi_0}U_\mu(x) & \textrm{for } \mu=1,2,3 \\
U_\mu(x) & \textrm{for } \mu=4\,,
\end{array}\right.
\end{equation}
and defining
\begin{equation}
\frac{1}{2\kappa} = \left(a_t m_q^0+\frac{3\nu}{\xi_0}+1\right).
\end{equation}
Thus we have $S_W=a_s^3\sum_x \bar\psi D\psi(x)$, where
\begin{equation}
D\psi(x) = \frac{1}{2\kappa}\psi(x) - \frac{1}{2}\sum_{\mu=1}^4 \left[
(1-\gamma_\mu)\tilde U_\mu(x)\psi(x+\hat\mu) +
(1+\gamma_\mu)\tilde U^\dagger_\mu(x-\hat\mu)\psi(x-\hat\mu)
\right]\,.
\end{equation}
(For convenience, we have also redefined $a_\mu\hat\mu \rightarrow \hat\mu$.)
Note that we have kept the continuum normalization of the fermion field
$\psi(x)$.

%%%%%%%%%%%%%%%%%%%%%%%%%%%%%%%%%%%%%%%%%%%%%%%%%%%%%%%%%%%%%%%%%%%%%%

\section{\label{quarkmass}Quark mass determination}

Due to the explicit breaking of chiral symmetry in the Wilson action,
the naive quark mass $m_q^0$ that appears in Eq.~(\ref{eq-sw-aniso}) is
not protected from additive shifts under renormalization.  In
Section~\ref{scalar}, we require the subtracted bare mass, $\widetilde m_s^0 = m_s^0
- m_\mathrm{crit}$, where the critical mass $m_\mathrm{crit}$
corresponds to the value of $m_0$ at which the physical quark mass
vanishes.  The naive mass for the strange quark is a parameter of the
theory; it has been chosen such that the mass of the $\phi$ meson
calculated on the lattice reproduces the physical
value~\cite{Bulava:2009jb}.  To determine $m_\mathrm{crit}$, we
utilize the dependence of the (partially quenched) pseudoscalar meson
mass $M_P$ on the valence quark mass: $M_P^2=2Bm_l^\mathrm{val}$ to
leading order in partially quenched chiral perturbation theory.  The
critical mass so defined depends implicitly on the fixed sea quark
mass~\cite{Eicker:1997ws,Sharpe:1997by}.  As advocated
in~\cite{Bhattacharya:1997ht}, a reasonable approach is to determine
$m_\mathrm{crit}$ for each available value of the light sea quark mass
$m_l^\mathrm{sea}$ and then extrapolate to the physical point where
$m_l^\mathrm{sea} = m_{u,d}$.  (Alternatively, one could fit all
available data for $M_P^2$ to a functional form that incorporates the
dependence on both the sea and valence quark
masses~\cite{Gockeler:2004rp}.)  Consistent with the other results
presented in this work, because only a single value of
$m_l^\mathrm{sea}$ is available, we do not perform this final
extrapolation in the light sea quark mass.

\begin{figure}
\begin{center}
\includegraphics[width=10cm]{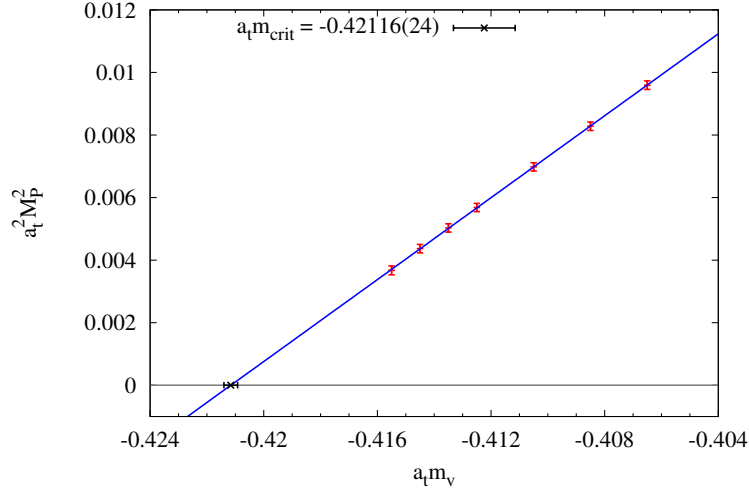}
\caption{\label{fig-mcrit}Mass of the pseudoscalar meson squared,
as a function of the valence quark mass.}
\end{center}
\end{figure}

Figure~\ref{fig-mcrit} illustrates our determination of the critical
mass.  For each of seven valence quark masses, we evaluated
pseudoscalar correlators on an ensemble of 216 gauge configurations.
The corresponding meson masses were determined from a single-cosh fit
in the range $20 \le t/a_t \le 44$.  Upon performing a linear
extrapolation in the quark mass, we find $a_t
m_\mathrm{crit}=-0.42116(24)$, where the statistical error has been
estimated via jackknife.  Given the naive mass that was input for the
strange quark, $a_t m_s^0=-0.38922$, we find $a_t \widetilde
m_s^0=0.03194(24)$ for the subtracted bare strange quark mass.

\section{\label{mixing} Flavor mixing for Wilson quarks}

In principle to extract continuum quantities, one must take the
lattice spacing (i.e., the bare coupling) to zero holding renormalized
parameters fixed, even for renormalization group invariant quantities
such as ratios of masses.  However, better estimates can often be found
at finite lattice spacing by ``renormalizing'' the bare lattice
quantities. A particularly interesting and difficult quantity for
Wilson fermions is the continuum parameter for the Higgs coupling to
the strange quark content of the nucleon,
\begin{equation}
f_{Ts} = \frac{m_s \langle N|\bar s s|N\rangle}{M_N} = 
m_s \frac{\partial}{\partial m_s} \log[ M_N]\,.
\label{eq-fts}
\end{equation}
The expression on the right is an identity based on the
Feynman-Hellmann theorem with the partial derivative taken with
respect to the renormalized strange quark mass, holding the
renormalized light quark mass and scale fixed.  For Wilson quarks on
the lattice, mixing with the light quark condensates in the nucleon
can produce a large contribution to $f_{Ts}$, as pointed out by
Michael, McNeile, and Hepburn~\cite{Michael:2001bv}. To understand
this, let us consider a mass-independent renormalization scheme on the
lattice.  We note that very similar methods are employed in Section
IV.B of \cite{Bhattacharya:2005rb}, despite some differences in the
choice of the lattice renormalization scheme. For convenience, in this
Appendix, we adopt a convention where lattice mass parameters are
dimensionless.
Also we will consider only the more physically relevant case of $2+1$ flavors,
even though our present results are calculated on 2 flavor gauge configurations.
%The mixing calculated here should still be present in the 2 flavor case.
There is however still non-zero mixing in the 2 flavor case as
illustrated in Figure~\ref{fig-mixing} to lowest order.

For Wilson quarks there are two important issues not present for a
chiral formulation.  First, we have additive mass renormalization,
which requires that a constant be subtracted from the bare quark mass.
Second, the disconnected diagram for the mass insertion operator
$\bar\psi \psi = \bar\psi_L \psi_R + \bar\psi_R \psi_L$ does not
vanish even when the subtracted quark masses vanish. We begin by
rewriting the mass term in the lattice Lagrangian in terms of singlet,
$m^0_S = (2 m^0_l + m^0_s)/3$, and non-singlet, $m^0_{NS} = (m^0_l -
m^0_s)/\sqrt{3}$, masses:
\begin{equation}
{\cal L}_m = m^0_l (\bar uu + \bar dd) + m^0_s \bar ss
 = m^0_S \bar\psi\psi + m^0_{NS} \bar\psi\lambda_8\psi\,,
\end{equation}
where for simplicity we take degenerate light quarks $m^0_l = m^0_u = m^0_d$.
The singlet and non-singlet masses renormalize differently because
of the lack of chiral symmetry.  The renormalization scheme 
we choose is
\begin{eqnarray}
m_S &=&  Z^m_0( g_0) (m^0_S  -m_\mathrm{crit}( g_0))/a\,, \nonumber\\
m_{NS}  &=&   Z^m_8 ( g_0) m^0_{NS}/a\,, \nonumber \\
\Lambda &=& \Lambda^0(g_0)/a\,.
\label{eq:RenParameters}
\end{eqnarray}
The first two equations define the renormalized masses, while the last
defines some renormalized scale which, for simplicity, we will take to
depend on the lattice spacing and some function of only the bare coupling.
There are many possible choices for $\Lambda$, such as the rho mass or
pion decay constant, though it will not enter into our final results so we 
leave it unspecified.
We have also introduced the lattice spacing, $a$, to
convert quantities to physical units.
This needs to be set by comparing some lattice measurement with a physical
value.
One choice would be to set the lattice spacing using the Sommer scale,
 $r_0$ \cite{Sommer:1993ce}, as
\begin{equation}
a = r_0^{phys} / r_0^{lat}(g_0)
\end{equation}
where $r_0^{lat}$ is the dimensionless value measured on each lattice ensemble
and $r_0^{phys}$ is some reference value in physical units.
As with $\Lambda$, the exact choice of definition for $a$ is irrelevant
for the present discussion.
Here we have made these quantities independent of the masses.
One could systematically improve on this by adding extra terms with
increasing powers of the subtracted masses, but, for simplicity,
we will not include these.

The mass renormalizations above can be expressed in
terms of the light and strange quarks themselves,
\begin{eqnarray}
m_l &=& m_S +  m_{NS}/\sqrt{3} = \frac{1}{3} \left[(2 Z^m_0  +
  Z^m_8)\widetilde m^0_l  + (Z^m_0 -
  Z^m_8)\widetilde m^0_s \right]/a\,, \nonumber \\
m_s  &=&  m_S - 2 m_{NS}/\sqrt{3} = \frac{1}{3} \left[(Z^m_0  + 2
  Z^m_8)\widetilde m^0_s   + 2 (Z^m_0 - Z^m_8)\widetilde
  m^0_l  \right]/a\,,
\label{eq:MassRen}
\end{eqnarray}
where $\widetilde m^0_i = m^0_i - m_\mathrm{crit}$.
We are now ready to find the expression for renormalized condensates.
Due to the vector Ward identity the non-singlet operator renormalizes as
\begin{equation}
(\bar \psi \lambda_8 \psi)^R = Z_8 (\bar \psi \lambda_8 \psi)^{lat}\,,
\end{equation}
where $Z_8 = 1/Z^m_8$, but the singlet piece is unconstrained by vector
current conservation.  However we can uniquely determine the
renormalization of the condensates by evaluating the Feynman-Hellmann
theorem,
\begin{equation}
f_{Ts} = m_s \frac{\partial}{\partial m_s} \log[ a^{-1} M^0_N]\,,
\end{equation}
(where $M^0_N = a M_N$ is the nucleon mass in lattice units)
in terms of bare lattice parameters.  The partial derivative is expanded as
\begin{equation}
\left.\frac{\partial}{\partial m_s}\right|_{m_l,\Lambda}=
\frac{\partial  m^0_s}{\partial  m_s} \frac{\partial}{\partial  m^0_s}
+ \frac{\partial  m^0_l}{\partial  m_s}  \frac{\partial}{\partial m^0_l}
+ \frac{\partial  g^{-2}_0}{\partial  m_s} \frac{\partial}{\partial
  g^{-2}_0 }\,,
\end{equation}
leading to the expression
\begin{equation}
f_{Ts} =  m_s \left[\; \frac{\partial  m^0_s}{\partial  m_s} \langle N|\bar s s|N\rangle_0 + \frac{\partial  m^0_l}{\partial  m_s}  \langle N|\bar u
u + \bar d d|N\rangle_0  + \frac{\partial  g^{-2}_0}{\partial  m_s}
 \left(\langle N|g_0^2 S_g|N\rangle_0  + a \frac{\partial a^{-1}}{\partial
   g^{-2}_0} M^0_N \right)\right]/M^0_N
\label{eq:Fexpansion}
\end{equation}
where $S_g$ is the gauge action and $\langle . \rangle_0$ is an unrenormalized
lattice matrix element.
%The last term appears to include a contribution proportional to
%$M^0_N$ (for $\langle N| \bar s s |N\rangle$)
% that should not be present in the operator product expansion.
%However we can show that it in fact contributes to and almost
%cancels the $\langle N|g_0^2 S_g|N\rangle_0$ term. The argument
%proceeds as follows. Consider defining a new ``lattice spacing''
%parameter, $\widetilde a = F(g_0, m^0_l, m^0_s)$, by fixing the
%nucleon mass $M_N$ instead of $\mu$.  The lattice spacing is
%fixed to obey the invariance condition
%\begin{equation}
%\widetilde a \frac{\partial }{\partial
%  g^{-2}_0} (\widetilde  a^{-1} M^0_N) = \langle N|g_0^2 S_g|N\rangle_0  -  g^3_0   M^0_N/2
%\beta_{N}(g_0, m^0_l, m^0_s) = 0 \,,
%\end{equation}
%where the first term uses the lattice Feynman-Hellmann theorem
%and the second term defines a new lattice beta function: $\beta_N(g_0,
%m^0_l, m^0_s) \equiv  - \widetilde 
%a \partial g_0/\partial \widetilde a $.  Substituting this expression
%for $M^0_N$ into Eq.~(\ref{eq:Fexpansion}), we obtain
%\begin{equation}
%f_{Ts} =  m_s \left[\; \frac{\partial  m^0_s}{\partial  m_s} \langle N|\bar s s|N\rangle_0 + \frac{\partial  m^0_l}{\partial  m_s}  \langle N|\bar u
%u + \bar d d|N\rangle_0  + \frac{\partial  g^{-2}_0}{\partial  m_s} c(g_0)
%  \langle N|g_0^2 S_g|N\rangle_0  \right]/M^0_N\,,
%\end{equation}
%the desired form, where $c(g_0) = 1- \beta_{N}(g_0)/\beta_{2}(g_0) \ne 0$ is
%$O(g^2_0)$ near the continuum limit because of the scheme independence
%of the two-loop beta function.

Finally, using the renormalization scheme in (\ref{eq:RenParameters}),
 we evaluate the coefficients in this
expression by use of the implicit function theorem,
inverting the Jacobian matrix
\begin{equation}
\frac{\partial(m_l,m_s, \Lambda)}{\partial(m^0_l,m^0_s,g^{-2}_0)}
= \left[
\begin{array}{ccc}
 \frac{\partial  m_l}{\partial  m^0_l} &  \frac{\partial  m_l}{\partial  m^0_s} & \frac{\partial m_l}{\partial g^{-2}_0}\\
 \frac{\partial  m_s}{\partial  m^0_l} &  \frac{\partial  m_s}{\partial  m^0_s}  & \frac{\partial m_s}{\partial g^{-2}_0} \\ 
0 & 0  &   \frac{\partial  \Lambda}{\partial  g^{-2}_0}
\end{array}
\right]\,.
\end{equation}
From Eq.~(\ref{eq:MassRen}) the determinant is then given by
\begin{equation}
J = \frac{\partial  \Lambda}{\partial  g^{-2}_0} \left[ \frac{\partial  m_l}{\partial  m^0_l} \frac{\partial  m_s}{\partial  m^0_s}  -
\frac{\partial  m_l}{\partial  m^0_s} \frac{\partial  m_s}{\partial  m^0_l}\right]
= \frac{\partial \Lambda}{\partial g^{-2}_0} Z^m_0 Z^m_{8} / a^2\,.
\end{equation}
Matrix elements of the inverse of the Jacobian are thus given by
\begin{eqnarray}
\frac{\partial  m^0_s}{\partial  m_s}  &=& J^{-1} \frac{\partial  \Lambda}{\partial  g^{-2}_0}  \frac{\partial  m_l}{\partial  m^0_l}  =
 \frac{a}{3} \left[ \frac{1}{Z^m_0} + \frac{2}{Z^m_8} \right]\,,
 \nonumber \\
\frac{\partial  m^0_l}{\partial  m_s}  &=&  - J^{-1} \frac{\partial
   \Lambda}{\partial  g^{-2}_0}  \frac{\partial  m_l}{\partial  m^0_s}
= \frac{a}{3} \left[ \frac{1}{Z^m_0} - \frac{1}{Z^m_8} \right]\,,
 \nonumber \\
\frac{\partial  g^{-2}_0}{\partial  m_s} &=& 0\,.
\end{eqnarray}
Thus we identify  $Z_0 = 1/Z^m_0$ and $Z_8 = 1/Z^m_8$ to obtain
 a form similar to Eq.~(\ref{eq:CondRen}) in the main text,
\begin{equation}
f_{Ts} =  \frac{m_s}{3 M_N} \left[\;( Z_0  + 2 Z_8)  \langle N|\bar s s|N\rangle_0 +
(Z_0 - Z_8) \langle N|\bar u u + \bar d d|N\rangle_0 \right]\,.
\end{equation}
%
%%% RCB
It is interesting that here the relations $Z_i = 1/Z^m_i$ did not
involve the use of Ward identities, contrary to standard derivations.
Also note that there is no $\langle N|g^2_0S_g|N\rangle_0$
contribution to this order in the renormalization scheme.
However as emphasized in 
\cite{Bhattacharya:2005rb}, it may be important to include
additional $O(a m_s)$ corrections which will cause $\partial
g^{-2}_0/\partial m_s$ to no longer vanish and will induce operator
mixing with $\langle N|S_g|N\rangle_0$. Since on dimensional and RG
grounds this term is $O(a m_s g^2_0)$, it should be relatively small.
The mixing with the light valence quarks is substantial and with the
estimate for $Z_8/Z_0 > 1$, it will tend to cancel the contribution
from $\langle N| \bar s s |N \rangle_0$ found for the bare amplitude.
To see this in more detail, note that renormalizing $f_{Ts}$ also
requires finding the renormalized strange quark mass,
\begin{equation}
m_s  = \frac{1}{3} \left[(Z^m_0  + 2 Z^m_8) \widetilde m^0_s+ 2 (Z^m_0
  - Z^m_8) \widetilde m^0_l \right]/a\,.
\end{equation}
Consequently, the dimensionless
ratio $f_{T_s}$ only depends on the  ratio
$Z^m_0/Z^m_8 = Z_8/Z_0$, which can be computed
by a procedure described in detail in~\cite{Rakow:2004vj,Gockeler:2004rp}
 in the context of quark masses.
It is convenient to rewrite the expression as 
\begin{equation}
f_{Ts}  =  \frac{ \widetilde m^0_s + 2 \Delta(\widetilde m^0_l - \widetilde m^0_s)/[3(1+\Delta)] }{M^0_N}
 \left[ \langle N|\bar s s|N\rangle_0 - \Delta 
\langle N|(\bar u u + \bar d d - 2 \bar s s)|N\rangle_0/3 \right]\,,
\end{equation}
in terms of an operator mixing parameter $\Delta = Z_8/Z_0 -1$.
The mixing term is a pure
non-singlet operator, $ \Delta \bar \psi \lambda_8 \psi/\sqrt{3}$.
The disconnected contribution to $ \bar \psi \lambda_8
\psi/\sqrt{3}$  vanishes like $O(a m_i)$ in the chiral limit,
but the valence contribution  remains large  at current lattice
spacings, resulting in a correction of the same order of magnitude as the
bare matrix
element given in the text: $a^{-1} \widetilde  m^0_s \langle N| \bar s
s|N\rangle_0 \simeq 504(91)(30)$~MeV.

\bibliography{strange}
\end{document}